\documentclass[prl,preprint,showpacs,byrevtex,endfloats,citeautoscript]{revtex4-1}
\usepackage{graphicx}
\usepackage{epsfig}
\usepackage{color}
\usepackage{xcolor}
\usepackage{amssymb}
\usepackage{amsmath}
\usepackage{amstext}
\begin{document}
\def\BibTeX{\rm B{\sc ib}\TeX}

\newcommand{\red}[1]{{\color{red}#1}}
\newcommand{\blue}[1]{{\color{blue}#1}}

\preprint{Simon et al.}

\title{Flat band and Lifschitz transition in long-range ordered supergraphene obtained by Erbium intercalation\\}

\author{A. Zaarour$^{1}$, V. Malesys$^{1}$, J. Teyssandier$^{1}$, M. Cranney$^{1}$, E. Denys$^{1}$,  J. L. Bubendorff$^{1}$, A. Florentin$^{1}$, L. Josien$^{1}$, F. Vonau$^{1}$,  D. Aubel$^{1}$, A. Ouerghi$^{2}$, C. Bena$^{3}$ and L. Simon$^{1}$ \footnote{corresponding author
\\E-mail address: Laurent.simon@uha.fr}}
\affiliation{$^{1}$Universit\'e de Haute-Alsace, CNRS, Institut de Science des Materiaux de Mulhouse UMR 7361, 3Bis rue Alfred Werner, 68093 Mulhouse, France. Universit\'e de Strasbourg, France}
\affiliation{$^{2}$Université Paris-Saclay, CNRS, Centre de Nanoscience et de Nanotechnologies, 91120, Palaiseau, France}
\affiliation{$^{3}$Institut de Physique Th\'eorique, Universit\'e Paris Saclay, CEA CNRS, Orme des Merisiers, 91190 Gif-sur-Yvette Cedex, France}

\date{\today}

\begin{abstract}

Dispersionless energy bands are a peculiar property gathering increasing attention for the emergence of novel photonic, magnetic and electronic properties. Here we report the first observation of a graphene superstructure n-doped up to the Lifshitz transition and exhibiting a flat band, obtained by ordered Erbium intercalation between a single layer graphene and SiC(0001). STM experiments reveal large graphene areas characterized by a long-range ordered hexagonal superstructure with a lattice parameter of 1.40 nm, rotated by $19^\circ$  with respect to the original lattice. Angle Resolved Photoelectron Spectroscopy measurements show that this graphene structure exhibits Dirac cones with perfect linear dispersion, and a Dirac point at $-1.72 eV \pm 0.02$  under the Fermi level, which is one of the highest doping levels ever obtained solely by intercalation. Fermi surface measurements show that the Lifshitz transition has been reached, and that a wide flat band is generated around the M point. We propose that this modification of the band structure is the effect of an induced spin-orbit coupling. This system provides a playground to study the interaction between a novel magnetic order mediated by $\pi$-band states, and a divergent density of states at the Fermi level.

\end{abstract}


\maketitle 
	Since the discovery of graphene, many efforts have been devoted to its functionalization and to the characterization of the resulting physical properties, thus paving the way for creating an increasing number of novel two-dimensional (2D) layered inorganic materials, and their combination in various 0D, 1D, and 2D van der Waals heterostructures \cite{JariwalaNatMat2017}. Doping by intercalation of metallic atoms was considered an efficient way to engineer the graphene electronic band structure. Earlier observations of intercalation involving graphene were obtained with noble metals such as gold \cite{PremlalAPL09,Cranney2010, NairPRB12,NairPRB16,FortiNatMat20}. Strongly inspired by the historical results observed in graphite intercalation compounds (GICs), and motivated by the possible observation of superconductivity for such systems, graphene became the investigation playground for intercalating a large variety of atoms such as Li, K, Ca, Cs and Rb \cite{SimonPSS18}. 

Among the numerous elements of the periodic table which have been studied for graphene intercalation, the Lanthanide atoms appear to provide higher levels of electron doping compared to alkali atoms. It became thus very interesting to explore the intercalated lanthanides, to see how they can dope the graphene, and also to test the interplay between the atomic magnetic moments and graphene, the possibility to induce spin-orbit coupling, and study the modifications of the graphene band structure.
		
It was also found that rare earth elements, and even aluminium and some of the above-mentioned alkali metals are highly reactive with graphene and can penetrate through several layers. A monolayer graphene (ML-G) on SiC(0001) consists in a buffer layer (BuL) graphene partially covalently bonded to the SiC substrate and of a graphene layer in Van der Waals interaction with the BuL. As it was first shown for hydrogen intercalation \cite{RiedlPRL2018}, by breaking the covalent bonds with the substrate, the BuL itself can be transformed in a decoupled graphene ML, which is denoted a quasi-free-standing monolayer graphene.

With the intercalation of Terbium atoms, by changing the ratio of BuL and ML-G on the pristine graphene prior to intercalation, some of us have definitively proved that the highly n-doped Dirac cone corresponds to the transformation of the buffer layer into monolayer graphene when the Tb atoms are intercalated between this layer and the substrate. We also have shown the possibility to n-dope graphene up to $4.3\times10^{14}cm^{-2}$, for which an electron-phonon coupling strength $\lambda$ of 1.2 was extracted from ARPES measurements, a value which is one order of magnitude higher than those usually observed in the alkali metals \cite{DaukyaPRB2018}. 

The highest value of n-doping observed so far has been obtained more recently by P. Rosenzweig et. al. \cite{StarkePRL2020} through a combination of Yb intercalation and a subsequent K adsorption on top. The Lifshitz transition, which corresponds to a graphene doping up to the point in which the Fermi level lies exactly at the Van Hove singularity, has been reached in this system.  The Fermi surface in this case shows a flat band around the M point, extending in the direction $\overline{KMK'}$ when doping by Yb intercalation, and becoming less and less elongated beyond the Lifshitz transition in the presence of K adsorption.  
The shape of the Fermi contour could certainly be perturbed by the hybridization of the Yb4f7/2 states and the $\pi$ band states of graphene as well as by the chemical disorder introduced by the deposition of K atoms on top of the n-doped ML graphene

More recently, by using the intercalation of Li atoms on G-ML on SiC(0001), a doubling of the Dirac cone has been observed, with the highly n-doped one corresponding to an electron density of $5.6 \times 10^{14} cm^{-2}$ \cite{BaoPRB2022}. In spite of this high electron density, the Lifshitz transition has not been reached. A so-called Van Hove singularity extension has been noted, with small flat band segments observed along the line connecting the apex of the two trigonal constant energy contours in the direction $\overline {KMK'}$. With Li atoms no hybridization with the graphene $\pi$ band state, nor any other interaction is expected, and the Fermi surface as a function of electron doping follows the same shape as expected for pristine graphene. For this system, a gap opening and a Kekul\'e-type order, evidenced by a $\sqrt{3}\times\sqrt{3} R 30^\circ$ reconstruction\cite{BaoPRL2021}, has also been observed. 

Beyond the classical BCS theory for the superconductivity in GiCs, with the recent discovery of the superconductivity in twisted bilayer graphene at the magic angle, and in rhombohedral graphene, all involving flat bands\cite{CaoNature2018paper1,CaoNature2018paper2,ZhouNature2021}, as well as with the systematic observation of flat bands in some high-Tc cuprate superconductors \cite{GofronPRL1994, LuPRL96, MazinPRL2005,IrkhinPRL2002}, the study of the shape of the Fermi surface, in particular the possibility to reach the flat bands around the M points in monolayer graphene is of high importance for many possible applications to realize non-conventional high-Tc superconductivity in the so-called Van Hove scenario \cite{NandkishoreNatPhys2012}. This is now becoming more realistic given the recent discovery of the possibility to realize a highly n-doped ML graphene up to the Lifshitz transition\cite{StarkePRL2020}.

Following this line of thought, we wanted to explore the properties of graphene in the presence of intercalated Er atoms. By using cycles of Erbium deposition and annealing, we have obtained a new type of graphene layer, fully developed on the surface, and apparent first in a $5.75\times5.75 R 19^\circ$ LEED-pattern with two domain orientations. The STM atomic-resolution images show a typical graphene layer, as well as a superposed structure with a periodicity of 1.40 nm turned by $19^\circ$ with respect to the graphene lattice. ARPES and XPS measurements show that the graphene layer is highly n-doped with a measured Fermi surface exhibiting a large flat band at the M point, and an electron doping level around the Lifshitz transition. The dispersion around the K point shows a perfect linearity up to the Fermi level, without renormalization such as a kink that could be attributed to electron-phonon coupling. Using tight-binding (TB) and T-matrix calculations we discuss the possible effect of a spin-orbit coupling induced by the magnetic Erbium atoms on the Fermi surface, which could explain the observed modifications in the band structure. While the observed superstructure is not fully understood in the framework of this paper, we tentatively explain it as a long range quasi-particle interference (QPI) pattern which is strongly affected by the modifications in the shape of the Fermi surface. This QPI pattern is possibly reinforced by long-range Kekul\'e order which arises not as a result of the intervalley scattering between momenta around the $\overline{K-K'}$ points of the Brillouin zone, as usually observed in the case of undoped graphene \cite{GutierrezNatPhys2016,CheianovSSC2009, XiaoPRB2018} but as the result of the peculiar Fermi surface structure at the Lifshitz transition.

\begin{figure*}
\includegraphics[width=17cm]{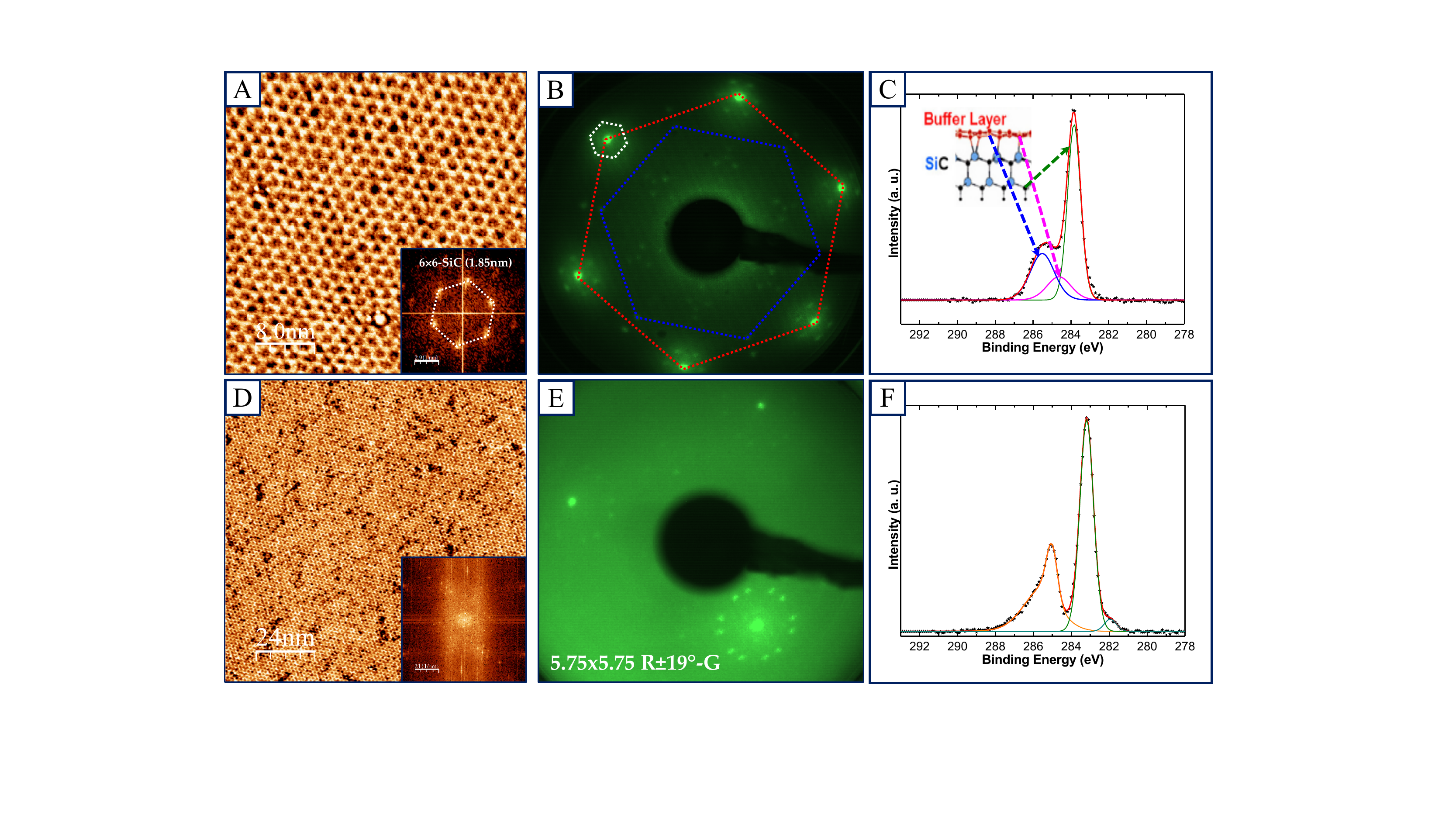}\\
\caption{STM image of the pristine Buffer graphene layer. A)$40\times40 nm^{2}, V_{S}=-1.67V, I=500pA$. The inset in A) shows the FT of the topographic image with the $6\times6$-SiC reconstruction which is found in the LEED pattern in B).  C) XPS C1s core level peak measured with a monochromatized $Al_{K_{\alpha}}$ source showing typical components attributed to the atoms indicated in the inset scheme.  STM topographic image after the ordered Er intercalation D) $-1.63 V; 500pA; 120\times120nm^2$ with the FT of the topographic image in the inset showing a monodomain. In E) the typical LEED pattern  around the specular spot shows here the $5.75\times5.75 R 19^\circ$ with two domain orientations. F) shows the typical C1s core level peak after Er intercalation under the BuL, with the disappearance of the components associated with BuL in favor of the component associated with graphene exhibiting an unusual strong asymmetry towards the largest binding energies which is attributed to the strong electron doping as discussed in the text (image processing using the WSxM software \cite{wsxm})}\label{Fig1}
\end{figure*}

In the present study, before intercalation we start with a well developed $(6\sqrt{3}\times 6\sqrt{3}) R30^\circ$ starting from the SiC(0001) substrate. The figure \ref{Fig1} A) shows a large area of buffer layer zoomed in B). The characteristic LEED pattern in C) shows the well-known $6\times6-SiC$ reconstruction relative to the $1\times1$ Bravais lattice of the topmost silicon atoms of the SiC (0001) substrate. The $C1s$ core level peak characteristic to a pure buffer layer is shown in D). As schematized in the inset, this core level peak is decomposed in three components, one corresponding to the carbon in the SiC substrate, one associated to the C-C bonds in the BuL, and one with the covalent bond between the C atoms in the BuL and the silicon of the substrate. To be sure to obtain the largest domain of pristine BuL, by minimizing the formation of monolayer graphene with respect to BuL, we have adjusted the temperature and time of annealing and we have followed the transition between the $5\times5$ reconstruction and the $6\times6$ one. 

The intercalation is performed by the deposition of 1ML of Erbium atoms following by an annealing at $1000^\circ$C. It is possible, by adjusting the annealing time and the deposition of Erbium (sometimes more than one Er ML is necessary, as a fraction of the Er is removed during the annealing process), to obtain the well-developed new structure depicted in figure \ref{Fig1} B). The STM image in E) shows large terraces, and the whole sample gives the characteristic LEED pattern in G), for which we show here the central specular spot. We observe a hexagonal structure giving two spots at a $\pm 19^\circ$ angle with respect to the $(1\times1)$ graphene lattice. These spots correspond to two domains which can be observed on the same terrace. Here the large terrace in E) corresponds to a mono domain. A zoom-in of the area in the red square is given in F) with a corresponding Fourier transform of the image in the inset. The core level peak C1s depicted in H) shows the component corresponding to the carbon in SiC, the disappearance of the two components corresponding to the BuL, and the formation of the new peak is attributed to the graphene layer. Usually, for a pristine ML graphene this component is much more symmetrical. Here it is necessary to add a supplementary component in the fitting procedure as this shows a peak tailing, characteristic of metallic systems but with a much higher asymmetry which is not usually observed. Such asymmetric core level peak has been attributed to a high level of electron doping \cite{SerneliusPRB2015}.

\begin{figure*}
\includegraphics[width=17cm]{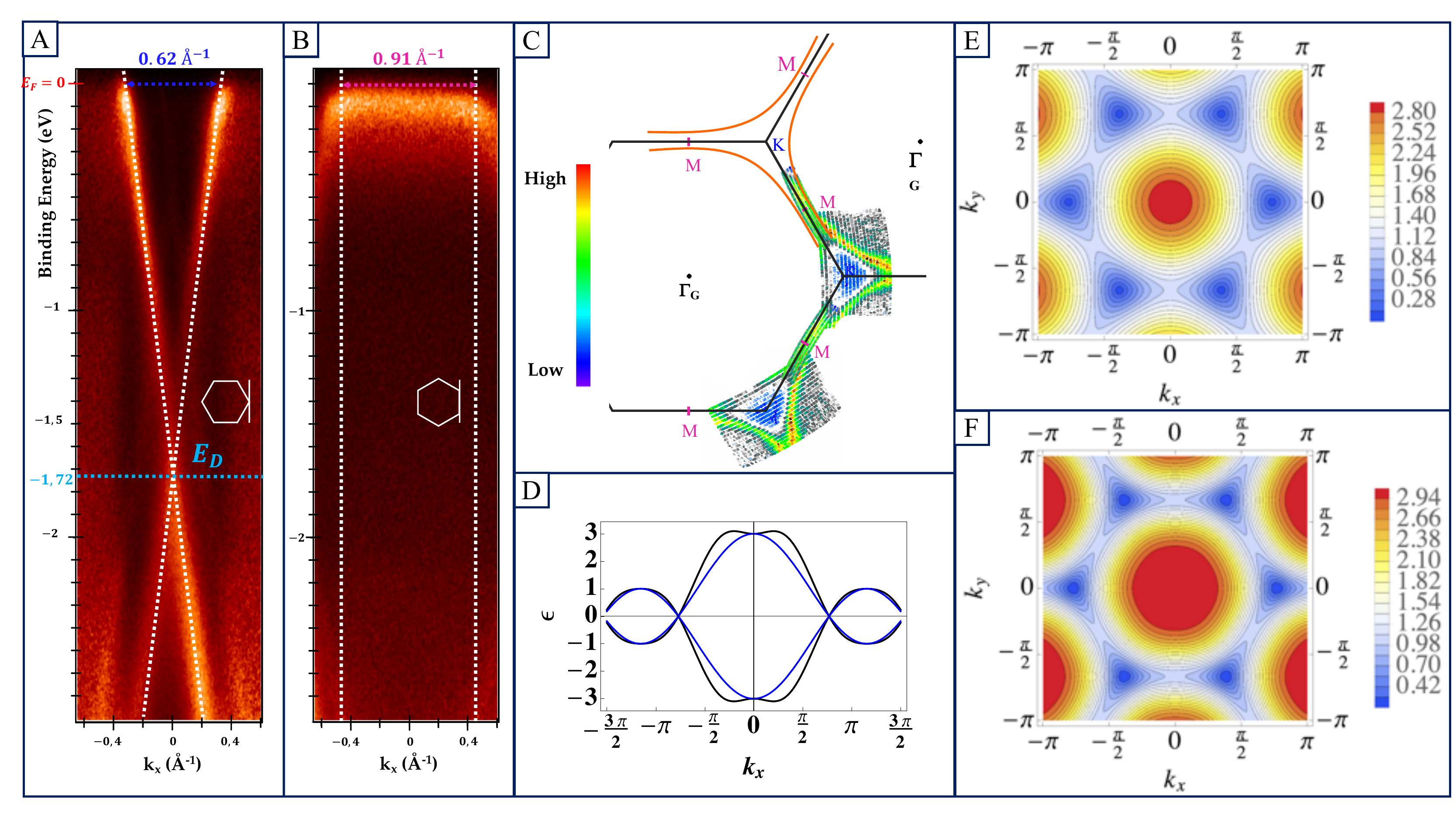}\\
\caption{ARPES measurements using an electron analyser Scienta-R3000 and a non-monochromatized UV sourve $h\nu=40.8eV$. The dispersion is measured around the K point in A), along the direction indicated in the inset, and around the M point in B). The Fermi surface is reconstructed and shown in C). D) Equal energy contours for pristine graphene with no spin-orbit coupling. E) An energy dispersion cut at $k_y=0$ for pristine graphene with no spin-orbit coupling (blue), and with an induced PIA of SOC of 0.8 (black), measured in units of the bandwidth (3.3eV). F) Equal energy contours for graphene with a 0.8 PIA spin-orbit coupling. }\label{Fig2}
\end{figure*}

Subsequently, in order to study the band structure of this system, we have performed Angle Resolved Photoemission-Spectroscopy (ARPES) measurements. This was done using an analyzer ScientaR3000 and a non-monochromatized UV source ($h\nu=40.8 eV$). The Figures \ref{Fig2} A), B) and C) show respectively the dispersion curves around the K and M points, and the Fermi surface. The inset schematics show the investigated paths in momentum space. Thus, in Fig. \ref{Fig2} A), in the vicinity of the K point we can see a Dirac cone with a perfect linear dispersion, and with a Dirac point at $-1.72$ eV under the Fermi level, without any gap opening or other modification of the dispersion.  As opposed to what we have previously observed for Tb intercalation \cite{DaukiyaPRB18} and in spite of the large n-doping, the dispersion remains linear up to the Fermi level, and no kink usually attributed to electron-phonon coupling is observed in this case in this dirrection. Around the point M, as depicted in Fig. ~\ref{Fig2} B), we observe a flat band, up to $0.91 \AA ^{-1}$ wide. The kink near the Fermi level is much more visible at the edge of the flat band than around the K point. 

Based on these observations of the dispersions around the K and M points, we have first deduced that the only possible Fermi surface contour is the orange line drawn in Fig. ~\ref{Fig2} C). This has been cross-checked with the measured Fermi surface, which is done by measuring the band dispersion while varying the angle $\theta$ and thus allowing the reconstruction of a part of the Fermi surface around the K and M points. We plot the results of these measurements at the scale of the first Brillouin zone in Fig. ~\ref{Fig2} C). We clearly see a triangular contour around the K point and the continuous flat band crossing the Van Hove singularity at the M point. We deduce that the Lifshitz transition is reached. The density of electrons calculated using the Luttinger theorem \cite{Luttinger} corresponds to the fillings of the region between the FS inner contour and the hexagon defining the first Brillouin zone is $5.1 \times 10^{14} cm^{-2}$, which is one of the highest electron doping ever observed solely by using intercalation, without any dopant deposited on top of graphene, as it was done using additional K in the case of Yb intercalation \cite{StarkePRL2020}. We think that this level of doping can be more easily reached with Erbium than Yb as the hybridization between the 4f states and the $\pi$ graphene band states has to be ruled out since the 4f states are at a lower binding energy than for Yb  (3 eV for Yb and 9 eV for Er) \cite{HandbookXPS}. We also notice that, in contrast to what has been pointed out recently by M. Jugovac et. al. \cite{JugovacPRB2022} for Li-intercalated graphene on Cobalt, we do not observe a gap opening; this is a supplementary argument ruling out a possible effect of the hybridization on the Fermi surface shape at the Lifshitz transition. 

To understand these modifications of the band structure and of the Fermi surface we investigate first if there are any covalent bonds forming between Si and Erbium. As both Si2p core level peaks measured by XPS are unchanged before and after Erbium intercalation we argue that this is not the case. In the same way, the Er4d spectrum measured after intercalation is strongly similar with the spectrum recorded on a 2.5 nm thick Erbium layer (Supplementary Information). The metallic character of the atoms is thus preserved. 

Without an evidence to suggest the formation of Er-Si bonds, we turn towards an explanation based on the magnetic nature of the Er atoms and the possibility to induce a non-zero spin-orbit coupling (SOC) in the monolayer graphene in their presence. Inducing an effective non-zero spin-orbit coupling by adatoms has been discussed for example in \cite{ModelSOKochanPRB2017}. Various SOC terms are discussed, including the Rashba SOC and the intrinsic Kane-Mele-like one. However we are looking here for a SOC mediated by the $\pi$ states in graphene, which preserves the band structure features that we observe, particularly no gap opening and a flattening of the Fermi surface along the direction $\overline{KMK'}$. It turns out that the only SOC term that can satisfy these constraints is the spin-flipping next nearest neighbor SOC, denoted principal-plane mirror asymmetry induced SOC, or PIA \cite{ModelSOKochanPRB2017}. 

We have performed tight-bind calculations using such an effective SOC term with various possible values of the induced SOC.
In Fig.~\ref{Fig2} D) we plot the energy dispersion obtained from diagonalizing the TB Hamiltonian in \cite{ModelSOKochanPRB2017}, the blue curve corresponds to the pristine graphene and the black curve to a PIA SOC with a value of 0.8 (in units of the graphene bandwidth, $3.3eV$). The calculations show indeed a linear dispersion at the K point, an increase of the Fermi velocity along the direction $\overline{\Gamma K M}$, no gap opening at the Dirac points, and a flattening of the band around the M point. The Figs.~\ref{Fig2} E) and F) show the Fermi contours at different energies for the dispersions given in D). This exemplarily demonstrates that for a PIA SOC coupling of 0.8 the main characteristic of the ARPES measured band structure are reproduced, for example the fact that the Fermi surface is elongated along $\overline{KMK'}$ same as the one measured by ARPES (see figure \ref{Fig2}C). This supports our hypothesis that the effect of the Er intercalation is to induce a large effective SO PIA coupling term in the quasi-free-standing monolayer graphene.

\begin{figure*}
\includegraphics[width=17cm]{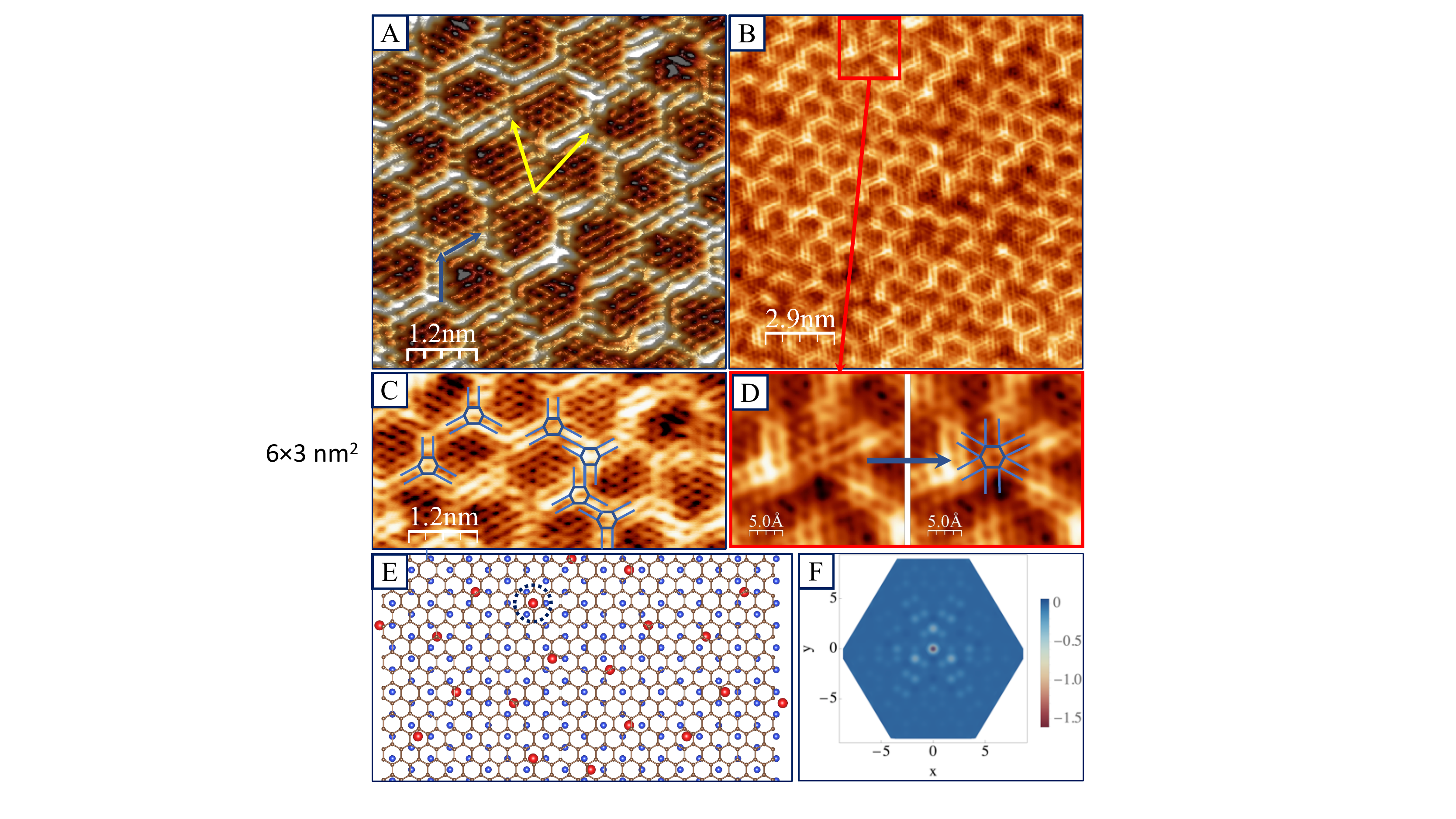}\\
\caption{Topographic STM images A) 3D representation of $6\times6 nm^2$; V=-20mV; I=500pA. Blue arrows show the direction of the graphene lattice, and the yellow arrows indicate the unit cell of the $5.75\times5.75 R 19^\circ$ superstructure. B)$15\times15 nm^2$; V=-207mV; I=500pA; In D) a zoom of the area in the red square shows a single impurity in the middle of a large hexagon with a 6-fold symmetry of the extended LDOS.  E) shows a possible model with a honeycomb organization of the intercalated Er atoms which act as impurities. The carbon atoms of the top graphene layer are denoted in black, the Er atoms in red, and the silicon atoms of the substrate in blue. The black dotted circle surrounding an Er atom at the $T_{4}$ site highlights the commensuration of the phase with the substrate. F) Modification of the LDOS in graphene induced at the Lifshitz transition by an impurity, for a SO PIA coupling of 0.8}\label{Fig3}
\end{figure*}

Let us now move to investigate the nature of this new phase as observed by STM measurements. 
In Fig.~\ref{Fig3} A) we show the 3D representation of a well-ordered region of the surface exhibiting the super-graphene phase. This 3D representation allows one to show the quasi-free-standing ML graphene which is fully continuous over the large hexagons. We see a higher intensity on the edges, and even something like an electron delocalization along the border of each large hexagon. As this is observed for different directions in the same image this is definitively not a tip artefact. In B) we depict a larger area with some defects in black, corresponding probably to missing intercalated atoms, as the top graphene layer is continuous over the entire area. 

This new structure is difficult to explain, particularly the angle of $\pm 19^\circ$ between the super-structure and the graphene lattice. First of all, the phase strongly looks like a Moir\'e pattern, and as depicted in supplementary information could be easily explained by having two superposed identical graphene layers rotated by $10^\circ$. However, as seen in the ARPES measurements we have only one graphene layer. We tentatively tried to explain this structure by other Moir\'e patterns, between the top graphene layer and an intercalated monolayer of Erbium, with or without the structure of the SiC(0001) underneath, by trying different lattice vectors for the Erbium ML, and different rotation angles between the atomic plans. Unfortunately, all these hypotheses failed to explain the structure. 

An accurate examination of this phase shows that the top graphene layer is perturbed along its lattice direction (see the blue arrows in Fig.~ \ref{Fig3} A). Thus, the only possibility to explain the observed ordered phase (defined by the lattice vector in yellow in Fig.~ \ref{Fig3} A)) and its disorientation with the graphene lattice, is to consider that we have Erbium intercalated atoms at the apex of a honeycomb structure with two atoms per unit cell and a lattice constant of 1.42nm, as schematized in Fig.~ \ref{Fig3}E). This first proposed simple model predicts the correct angle of $19^\circ$ with the original graphene lattice. We have found that this structure corresponds to a commensuration of Erbium atoms (in red) with the $T_4$ site of the topmost $1\times 1$ silicon atoms (in blue) of the SiC substrate (see the dotted circle in Fig.~ \ref{Fig3} E)). This is the only commensuration that we have noted. 

From XPS measurements we estimated at 2.8\% the ratio of Er atom respectively to the C atoms in the graphene layer \cite{XPS}. With this first model we have $3\%$ of Erbium and considering that we can have up to 3 electrons per each Er atom ($Er^{3+}$), the expected electron density is $3.5\times 10^{14}cm^{-2}$ which is lower but comparable with the density evaluated from ARPES. Both Erbium content and electron density are in accordance with the measurements. We notice that the precision of our Fermi surface measurement is quite low due to the thickness of our analyser slit.

Given the absence of modifications, and of formation of new covalent bonds, we argue that the observed graphene superstructure is rather explained by the effect of intercalated Erbium atoms under the top graphene layer acting as impurities which modify the local density of states of the graphene monolayer (LDOS). We base our argument on analyzing first an isolated defect in graphene (see Figs.~\ref{Fig3}B) and D)), which is not a part of the ordered Er superstructure. For such a defect we observe a small hexagon in the center, which corresponds to an aromatic carbon ring. The density of state seems to extend into a six-fold star-like structure with six wings originating on the six edges of the aromatic ring. We depict a schematics of this six-fold (6C) symmetry structure by the blue lines in Fig.~\ref{Fig3}D). 

We compare this to a real-space T-matrix formalism \cite{SIMONJPhysD} of the impurity-induced LDOS in the vicinity of a localized impurity in graphene at the Lifshitz transition, in the presence of the effective spin-orbit coupling (SOC PIA of 0.8) that we have proposed to explain the observed characteristics of the band structure measured by ARPES. The resulting QPI is depicted in Fig.~\ref{Fig3}F).  We see that indeed the theoretical calculation reproduces reasonably well the observed impurity feature.

When the Er impurities are ordered into a honeycomb structure it appears that the LDOS in the vicinity of each impurity exhibits rather an effective three-fold (3C) symmetry. The difference between the network 3C symmetry and single impurity 6C symmetries is discussed in what follows. By schematizing the impurities as small hexagons and a star-like feature made of three sets of double lines originating on three sides of the hexagon, symbolizing the extension of the LDOS modifications, we see that it is possible to generate the super-graphene phase (see Figs.~\ref{Fig3}C)). The angle of $\pm 19^\circ$ is thus a result of the position of the Er impurities which also gives rise to a register shift between the pair of lines which schematize the delocalization of charge. 

\begin{figure*}
\includegraphics[width=17cm]{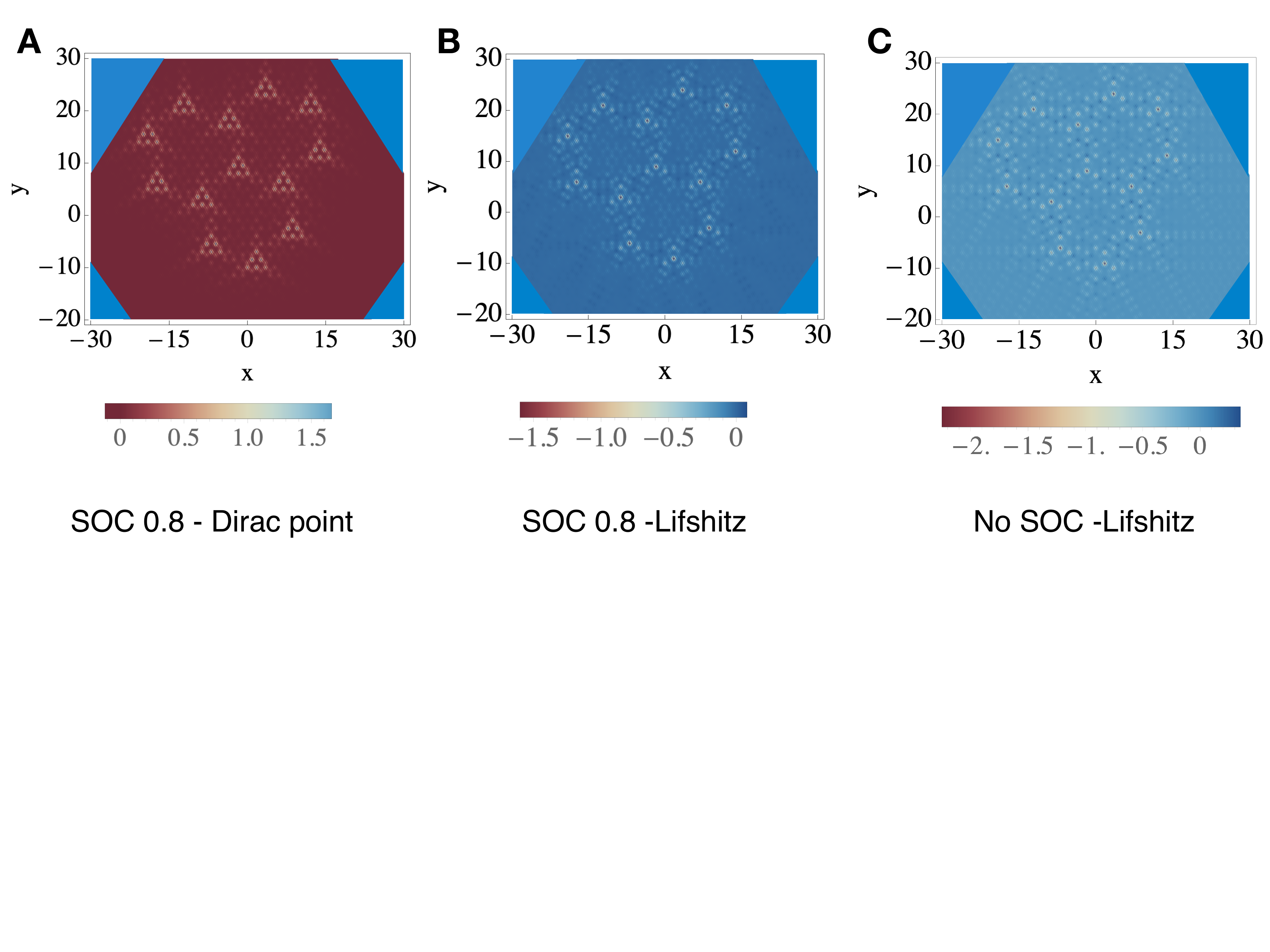}\\
\caption{Modifications induced by a network of impurities to the LDOS of graphene: A) at the Dirac-point energy with no spin-orbit coupling, B) at the Lifshitz transition with no-spin orbit coupling, and C) at the Lifshitz transition with a PIA spin-orbit coupling of 0.8. We take the impurity potential to be equal to 100 times the bandwidth but the results depend very little qualitatively on the chosen value of the impurity potential.}\label{Fig4}
\end{figure*}

This simple picture is supported by a detailed theoretical analysis. In Figs.~\ref{Fig4} A), B) and C), using the T-matrix formalism, we have calculated the effect of a number of impurities placed at the positions expected for the Er atoms in the hexagonal superstructure on the LDOS.  For comparison we consider the LDOS fluctuations induced in graphene with an effective PIA SOC of 0.8 at the Dirac point (A) and at the Lifshitz transition (B), as well as for a pristine graphene layer  at the Lifshitz transition (B). While we observe a 'standard' $\sqrt{3}\times\sqrt{3} R 30^\circ$ reconstruction around the defect for the LDOS at the Dirac point, we note that, consistent with the single impurity calculation in Fig.~\ref{Fig3}D), for the energy corresponding to the Lifshitz transition each impurity gives rise rather to a LDOS pattern with a six-fold symmetry, similarly to the STM patterns observed in Fig.~\ref{Fig3}. When ordering the Er impurities into a honeycomb lattice slightly rotated with respect to the graphene one, the effect is to enhance the visibility of the LDOS modifications along the three lines connecting one Er atom to his neighbors and thus give rise to an apparent 3C symmetry rather than to a 6C one as it is the case for an isolated impurity. Moreover, the induced PIA SOC also enhances the interference effect and the intensity of the standing wave along the direction which links the two impurities. This allows us to obtain a good agreement with the observed $5.75\times5.75 R 19^\circ$ phase for the LDOS in graphene.

There is another effect that may contribute to these observations that cannot be taken into account by our coupled TB and T-matrix calculation: it is highly likely that the observed super-graphene phase is also characterized by a Kekul\'e distortion (KD) which modifies even more deeply the QPI patterns, and in particular contributes to the reinforcement of the density of state interference between the intercalated Er atoms which act as impurities. Such KD order has been observed in the case of graphene on Ag(111) surface where vacancies on the substrate act as impurities for the top graphene layer. In the case of undoped graphene, the Fermi level is at the Dirac point, and the resulting standing wave pattern, i.e. the electronic ripples observed around the impurities/adatoms, will give rise to a $\sqrt{3}\times\sqrt{3} R 30^\circ$ reconstruction. When the distance and the orientation between the impurities is in commensuration with the standing wave pattern, the whole surface gives rise to a $\sqrt{3}\times\sqrt{3} R 30^\circ$ structure, and, as pointed out by Gutierez et. al. \cite{GutierezNatPhys2016}, when the ripples are in phase, the electron-lattice interaction gives rise to atomic displacements of the carbon atoms and to the development of the Kekul\'e distortion phase. 

For our system the Fermi level is at the Lifshitz transition and the QPI patterns are characterized by the structure described in Fig.~\ref{Fig3}D). Assuming that a KD associated with these patterns arises, we expect that some of the effects observed in the impurity-perturbed LDOS to be enhanced, as well as to be generalized at all energies: when a strong order appears because of the KD, the standing wave patterns are rather charge density waves (CDW) than Friedel oscillations (CDW differ from Friedel oscillation by the fact that there is no dispersion with energy). Indeed, we have measured the Fourier Transform of the $dI/dV$ map images and found that the maps are energy independent. Nevertheless, the STM image (in direct space) shows that the intensity of the edges of the hexagon is high at the Fermi level and decreases for higher energies.  This behavior is also certainly due to the LDOS being strongly pinned at the Fermi surface due to the flat band. However, a more detailed experimental and theoretical analysis would be necessary to argue definitely in the favor of the existence of a Kekul\'e distortion phase. Our results provide a starting point to investigate the QPI patterns, as well as the existence of a possible KD in graphene when the Fermi level lies at the Lifshitz transition.

In conclusion we report here for the first time a novel, ordered, highly n-doped, free-standing super-graphene phase obtained by the intercalation of Erbium atoms. The intercalated Erbium atoms seem to lie free in-between the top graphene layer and the SiC substrate, forming no bond with the silicon atoms. The band structure measured by ARPES reveals a robust linear dispersion with a Dirac point at an energy of -1.72 eV under the Fermi level, and no gap opening. The measurement of the Fermi surface indicates that the Lifshitz transition has been reached with an electron density reaching $5.1\pm 0.8 \times 10^{14} cm^{-2}$ and reveals a flattening of the band around the M point.  Using Tight-binding calculations we have studied the possibility to induce an effective SOC in graphene in this configuration, that can explain in particular the modifications of the band structure and the flattening of the Fermi contour around the M points. The observed hexagonal superstructure is in good agreement to impurity-induced features generated by the ordered  Er atom network, and raises questions about the existence of a Kekul\'e distortion phase. While we are aware that the model proposed here may not be the only explanation for the observed features, the present manuscript provides substantial new insight to motivate further experiments and theoretical analysis of this problem.

\acknowledgments 
This work was supported by the R\'{e}gion Grand Est and European fond Feder through the 'NanoteraHertz' project and by the French
National Research Agency (ANR) through the project MIXES (grant ANR-19-CE09-0028).\\

\newpage

\begin{thebibliography}{60}
\bibitem{XPS} The Er content is calculated with the ratio of the $4d_3/2$ core level component of Er 4d by the area of the C1s-G component of the C1s core level component. A RSF of 12.6 is considered giving 5.04 for the $4d_{3/2}$ component.
\end{thebibliography}
\end{document}